\documentclass[12pt]{article}
\usepackage[dvips]{color}
\usepackage{mathtext}
\usepackage[koi8-r]{inputenc}
\usepackage{graphicx}
\usepackage[figuresright]{rotating}
\usepackage{amssymb,eucal}
\usepackage{cite}

\newcommand{\beq}{\begin{equation}}
\newcommand{\eeq}{\end{equation}}
\newcommand{\beqn}{\begin{eqnarray}}
\newcommand{\eeqn}{\end{eqnarray}}

\date{}
\begin{document}
\title{Scintillator counters with multi-pixel avalanche 
photodiode  readout for the ND280 detector of the T2K experiment}
\author{O.~Mineev$^a$\thanks{Corresponding author.
{\it Email address:} oleg@inr.ru}, A.~Afanasjev$^a$, G.~Bondarenko$^b$,
V.~Golovin$^b$, E.~Gushchin$^a$,\\  
A.~Izmailov$^a$,  M.~Khabibullin$^a$, A.~Khotjantsev$^a$, Yu.~Kudenko$^a$, \\
Y.~Kurimoto$^c$, T.~Kutter$^d$, B.~Lubsandorzhiev$^a$, 
V.~Mayatski$^e$,  \\ 
Yu.~Musienko$^a$, T.~Nakaya$^c$,  T.~Nobuhara$^c$, B.A.J.~Shaibonov$^a$, \\
 A.~Shaikhiev$^a$,  M.~Taguchi$^c$,   N.~Yershov$^a$, M.~Yokoyama$^c$ \\ 
 {} \\
$^a${\it Institute for Nuclear Research RAS, 117312 Moscow, Russia} \\ 
$^b${\it Center of Perspective Technology and Apparatus,
107076 Moscow, Russia} \\
$^c${\it Department of Physics, Kyoto University, Kyoto 606-8502, Japan} \\
$^d${\it Department of Physics and Astronomy, Louisiana State University} \\
{\it Baton Rouge, Louisiana 70803-4001, USA} \\
$^e${\it AO Uniplast, 600016 Vladimir, Russia}}
\maketitle
\begin{abstract} 
The Tokai--to--Kamioka (T2K) experiment is a second generation 
long baseline neutrino oscillation experiment which  aims at a sensitive 
search for the $\nu_e$ appearance. The main design
features of the T2K near neutrino detectors located at 280 m from the target are 
presented. Scintillator counters  developed for the T2K near detectors 
are described. Readout of the counters is provided via WLS fibers embedded into S--shape grooves 
in a scintillator and viewed from both ends by  multi--pixel avalanche
photodiodes operating in a limited Geiger mode. A description, operational
principles and the results of  tests of photosensors with a sensitive area of 
1.1 mm$^2$ are presented. 
A time resolution of 1.5 ns, a spatial resolution of 9.9--12.4 cm, and 
a  MIP detection efficiency of  more than 99\% were obtained for scintillator 
detectors in a  beam test. 
\end{abstract}
~\\

\section{Introduction}
The T2K project~\cite{t2k} is a second generation long baseline neutrino 
oscillation
experiment which will use a high intensity  off--axis neutrino beam produced 
by the JPARC 50 GeV (initially 40 GeV) proton beam. The first phase of the
T2K experiment   pursues two main goals: a sensitive 
measurement of ${\rm\theta}_{13}$  and  a more accurate determination of the 
parameters ${\rm sin^22\theta}_{23}$ and $\Delta m^2_{23}$ than any previous
experiment.

\begin{figure}[htb]
\centering\includegraphics[width=11cm,angle=0]{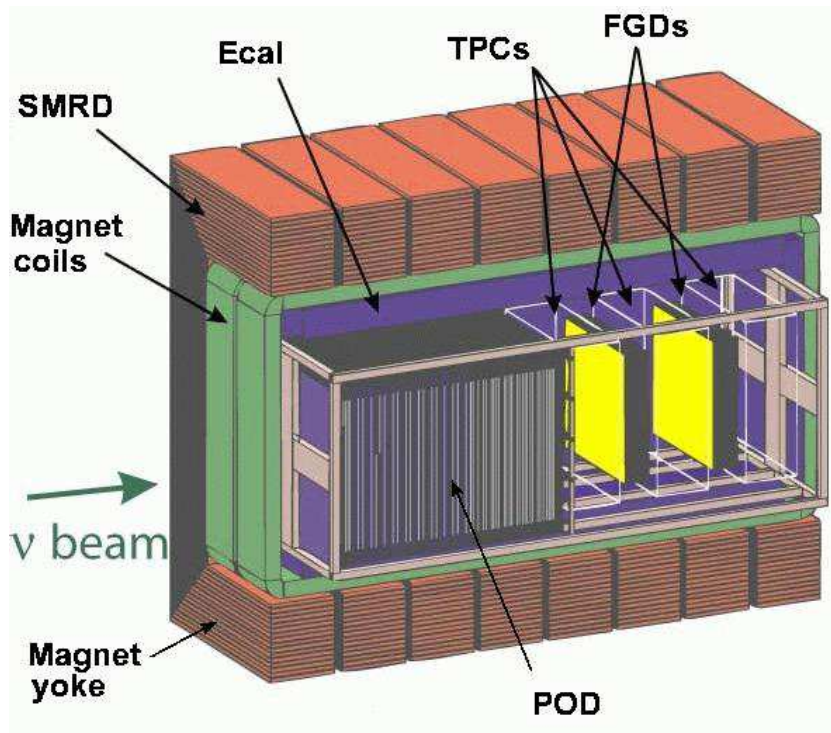}   
\caption{The cutaway view of the T2K  near detector.}
\label{fig:nd280}
\end{figure}

To achieve the physics goals, it is important to provide precise measurements 
of the neutrino beam properties, neutrino flux, spectrum and 
interaction cross sections. For these purposes, the near detector complex
(ND280~\cite{nd280})  will 
be built at the distance of 280 m from the target along the line between the average pion 
decay point and the Super-Kamiokande detector. In order to keep the systematics
uncertainties below the statistical error, the physics requirements for ND280, 
discussed in detail in Ref.~\cite{nd280}, can be briefly summarized as 
follows. The energy scale of the neutrino spectrum 
must
be understood at the 2\% level, and the neutrino flux should be monitored with 
better than 5\% accuracy. The  momentum resolution of muons from the 
charged current
quasi--elastic interactions should be less than 10\%, and the threshold for 
the
detection of the recoil protons  is required to be about 200~MeV/c. The 
$\nu_e$ fraction is to be measured with an uncertainty of better than 10\%.
The measurement of  the neutrino 
beam direction with precision much better than 1 mrad is expected to be 
provided by the
on--axis detector (neutrino monitor).
The off--axis ND280 is shown in Fig.~\ref{fig:nd280}
and consists of the UA1 magnet operated with a magnetic field of 0.2 T, 
a Pi-Zero detector (POD), a tracking detector which includes  time projection 
chambers (TPC's) and fine grained scintillator detectors (FGD's), an 
electromagnetic calorimeter
(Ecal), and a side muon range detector (SMRD).

 The POD has been designed to be similar to the  
 MINER$\nu$A detector~\cite{minerva}.
It is installed in the upstream end of the magnet and 
optimized for measurement of the inclusive  $\pi^0$ production by 
$\nu_{\mu}$ on oxygen.  The POD consists of 76 tracking planes  composed of 
triangular polystyrene  scintillating bars (about $2\times 10^4$) alternating 
with 
thin (0.6 mm) lead foils. Each bar has a 3 cm base, 1.5 cm height and 
a central  hole for a WLS fiber.  Three TPC's will measure the 3--momenta of 
muons produced by charged current interactions in the detector  and will 
provide the 
most accurate measurements of the neutrino energy spectrum. The ND280 will 
contain two FGD's, each 
with dimensions $2\times 2\times 0.3$ m$^3$ resulting in a  total target mass 
of 1.2 tons.  
The first FGD will be an active scintillator detector, similar to the 
SciBar~\cite{scibar} detector of the K2K experiment~\cite{k2k}.
Each FGD layer  will
consist of 200 scintillator bars, and thirty layers will be  arranged 
in alternating vertical and horizontal layers perpendicular to the beam
direction.   The second FGD will consists of 
$x - y$ layers scintillator bars alternating with 3 cm thick layers of passive
water.  The Ecal surrounds the POD and 
tracking region and consists of 15 layers, each
of which has one sheet of 3 mm Pb alloy and one layer of 1 cm thick $\times$ 
5 cm wide 
plastic scintillator bars (with signal readout by WLS fibers). Interior to this
section is the preradiator section, where each of the 3 layers consists of a
lead alloy sheet backed by 3 layers of scintillator bars. Air gaps in the 
UA1 magnet will be instrumented with plastic scintillator to measure the ranges
of muons which escape at large angles with respect to the neutrino 
beam and which can not be measured by the TPC's. The active component of the 
SMRD will use scintillators  with wavelength shifting fibers~(WLS) readout 
to transport the light into photosensors.

The ND280 detector will widely use WLS fiber readout with light
detection   from fibers by photosensors which have to operate 
in a magnetic field environment and  limited  space inside the UA1 magnet. The primary candidate
for the photosensor is the multi--pixel avalanche photo-diode operating in the 
limited Geiger multiplication mode~\cite{dev1,dev2,dev3,dev4}. Such photodiodes 
are compact, well matched to spectral emission of WLS fibers, and insensitive 
to magnetic fields~\cite{andreev,beznosko}.

\section{Geiger mode multi--pixel avalanche photodiodes }
\subsection{Overview}
\begin{figure}[htb]
\centering\includegraphics[width=15cm,angle=0]{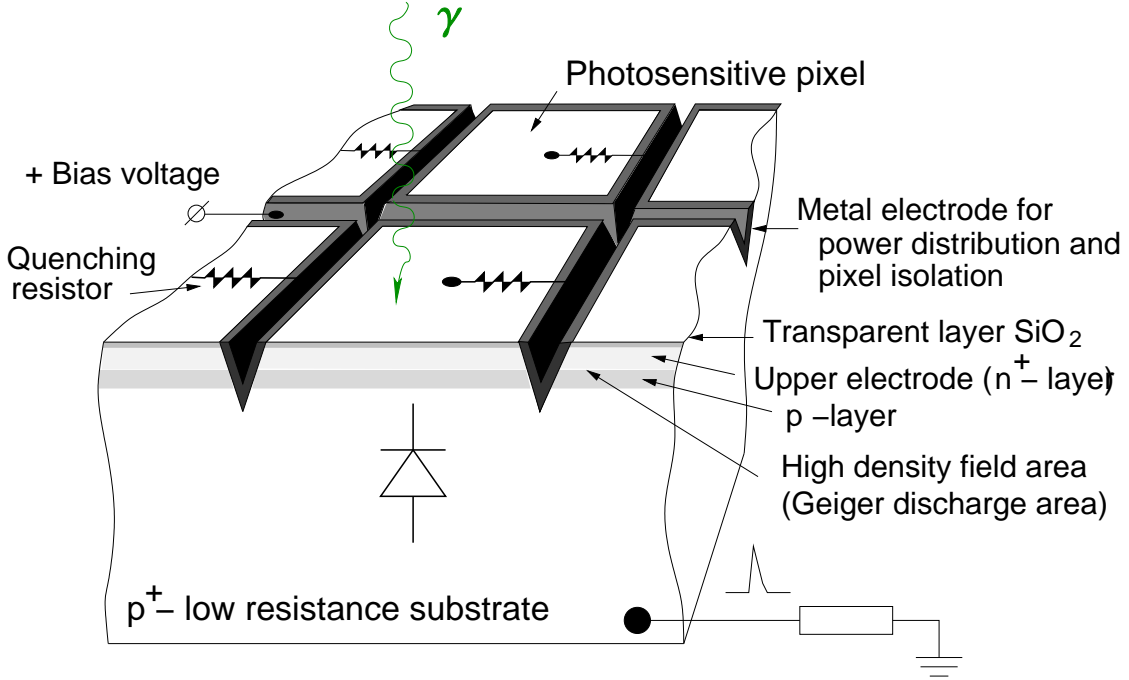}   
\caption{The schematic view of the MRS photodiode structure.}
\label{fig:scheme_mrs}
\end{figure}

The  
multi-pixel avalanche photodiodes  with a  metal--resistor--semiconductor 
layer  structure  operating in the limited Geiger mode 
(hereafter referred to as MRS APD's 
or MRS photodiodes) are invented and designed by the Center of Perspective
Technologies and Apparatus (CPTA), Moscow~\cite{dev1}. 
Detailed description of these devices and principles of operation 
 can be found in Refs.~\cite{dev-dop,dev3,si1}.
Such a  photosensor  consists of many pixels on a common p--type silicon 
substrate. A simplified topology of a MRS photodiode is shown in 
Fig.~\ref{fig:scheme_mrs}.
 Each pixel operates as an independent
Geiger micro-counter with a gain of the same order as a vacuum 
photomultiplier. Geiger discharge is initiated by a photoelectron in the 
high electric field created  in a very thin layer of about 
1~$\mu$m by the applied  bias voltage.  
The discharge current for each pixel produces a voltage drop at  
individual  resistors. As a result, the electric field density 
becomes small and can no longer  support the discharge quenched in such a way.
Small pixels are separated by grooves filled with an optically non--transparent 
material to suppress the cross--talks. The gain is determined by the charge 
accumulated in a pixel capacitance: $Q_{pixel} = C_{pixel}\cdot\Delta V$, 
where $\Delta V$ is difference between the bias voltage and the breakdown 
voltage of the diode (overvoltage). Since $\Delta V$ is about a few volts and 
$C_{pixel}\simeq 50$ fF, then typical  $Q_{pixel}\simeq 150$ fC, that 
corresponds to $10^6$ electrons.  A single incident photon can fire more 
than one pixel. Thus, the gain of the MRS photodiode is equal to the charge 
of a pixel multiplied by the average number of pixels fired by a 
single photon. 

\begin{figure}[htb]
\centering\includegraphics[width=14cm,angle=0]{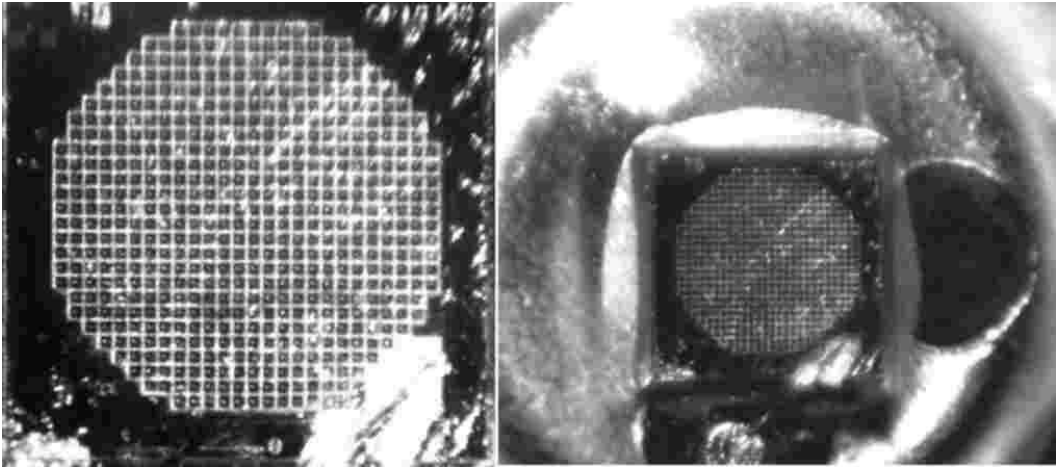}
\caption{Face view of a MRS avalanche photodiode with 556 pixels,  
magnified 56 times (left) and 14 times (right). The sensitive area has an 
octagonal shape with an 1.2 mm side--to--side distance.}
\label{fig:mrs} 
\end{figure}

The amplitude of a single pixel signal does not depend on the triggered 
number of
carriers in this pixel. In such a way, the photodiode signal is a sum of 
fired pixels. Each pixel operates as a binary device, but the multi--pixel 
photodiode as a  whole unit is 
an analogue detector 
with a dynamic range limited by the finite number of pixels.  
The pixel size can be $15\times 15$ to $70\times 70$~$\mu$m$^2$, and the total 
number of pixels 
is 100--4000 per mm$^2$. We tested the CPTA MRS photodiodes with a 
sensitive area of 1.1~mm$^2$ 
with 556 pixels of $45\times45~\mu {\rm m}^2$ size (see Fig.~\ref{fig:mrs}). 

\subsection{Basic properties of the CPTA MRS photodiodes}

The main parameters of the MRS photodiodes such as the  gain, photon detection 
efficiency, 
intrinsic noise, cross--talk, depend on the applied bias voltage. The MRS
photodiodes were tested using a signal from  a green light emitting diode 
(LED).  The MRS signal was amplified and split into two signals: 
one was fed to a discriminator, another was measured by an ADC with a 
gate of about 100 ns.

{\it Gain.}
A typical operating  voltage is around 40~V for the tested  MRS photodiodes, 
although the voltage can differ by 
a few volts to provide the same gain in photodiodes.
The MRS photodiode has an excellent single photoelectron~(p.e.) resolution 
determined mainly 
by electronics noise even at room temperature. It means that there is only
a small pixel 
to pixel gain variation as well as small fluctuations in Geiger discharge 
development. The  absolute gain 
depends on the photodiode topology, bias voltage and 
temperature. The voltage and temperature sensitivities of the MRS photodiode 
gain are rather weak as will 
be demonstrated  below.   A 0.1~V  change in 
bias voltage corresponds to a 2--3\% variation in gain.     The charge 
of a single p.e. signal in a calibrated ADC was used 
to determine the MRS photodiode gain.
The typical gain value   at room temperature 
(22$^{\circ}$C) is obtained 
to be about $0.5\times 10^6$. 

\begin{figure}[htb]
\centering\includegraphics[width=12cm,angle=0]{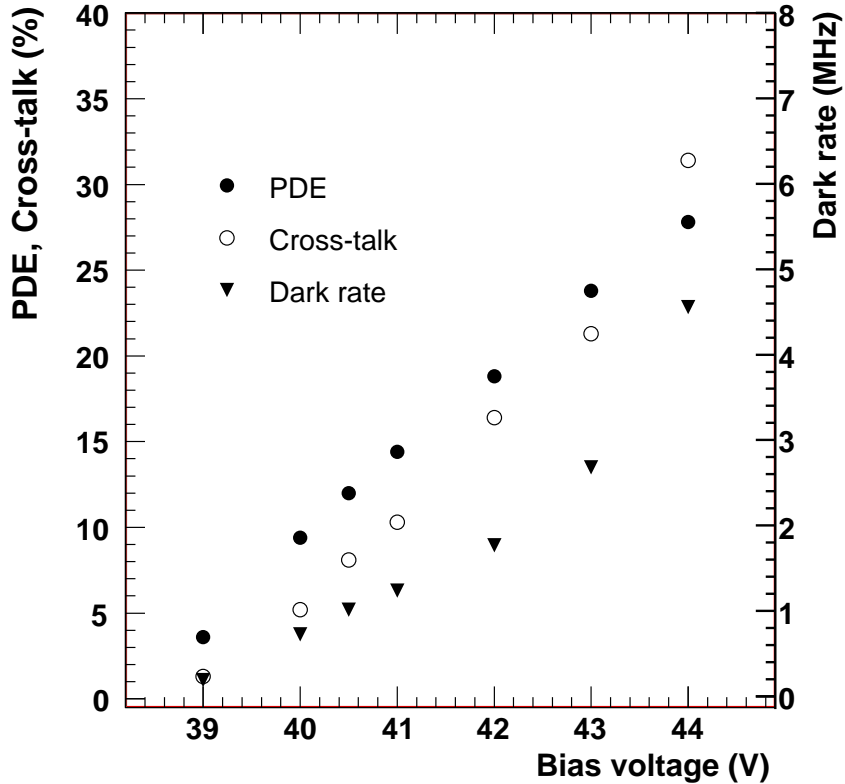}
\caption{The photon detection efficiency, cross--talk and dark rate 
as a function of the applied bias voltage.}
\label{fig:pde} 
\end{figure}

{\it Photon detection efficiency.} 
The  photon detection efficiency (PDE) of a multi-pixel avalanche photodiode 
operating in the limited Geiger mode  is a product of 3  factors:
\begin{equation}
{\rm PDE} = QE\cdot\varepsilon_{Geiger}\cdot\varepsilon_{pixel},
\label{eq:pde}
\end{equation}
 where $QE$ is the 
wavelength 
dependent quantum efficiency, 
$\varepsilon_{Geiger}$ is the probability for a photoelectron to initiate the 
Geiger discharge, 
$\varepsilon_{pixel}$ is a fraction of the total photodiode  area 
occupied by sensitive pixels.
 The  bias voltage affects 
one parameter  in expression~(\ref{eq:pde}) $\varepsilon_{Geiger}$.   
The geometrical factor $\varepsilon_{pixel}$ 
is  completely determined by the MRS photodiode topology. Its value is 
estimated to be about 
70--80\% in  an ideal case.  The absolute value of the PDE was measured using 
small pulses from a  green LED (525 nm) which illuminated a MRS photodiode 
through  a 0.5 mm diameter collimator. The number of
emitted photons was obtained using a calibrated PMT XP2020. 
The  normalized PDE  values at 525~nm  for different bias voltages are 
presented in  Fig.~\ref{fig:pde}.
 The cross-talk 
contribution was subtracted from the signal to obtain  the correct value 
of the PDE,  the accuracy of which is estimated to be  about 20\%.
As seen in  Fig.~\ref{fig:pde},  
the PDE is about 12\% at a dark rate of 1~MHz. 
The PDE can be increased up to almost 30\% at the expense of much higher dark
rate.   
 
\begin{figure}[htb]
\centering\includegraphics[width=12cm,angle=0]{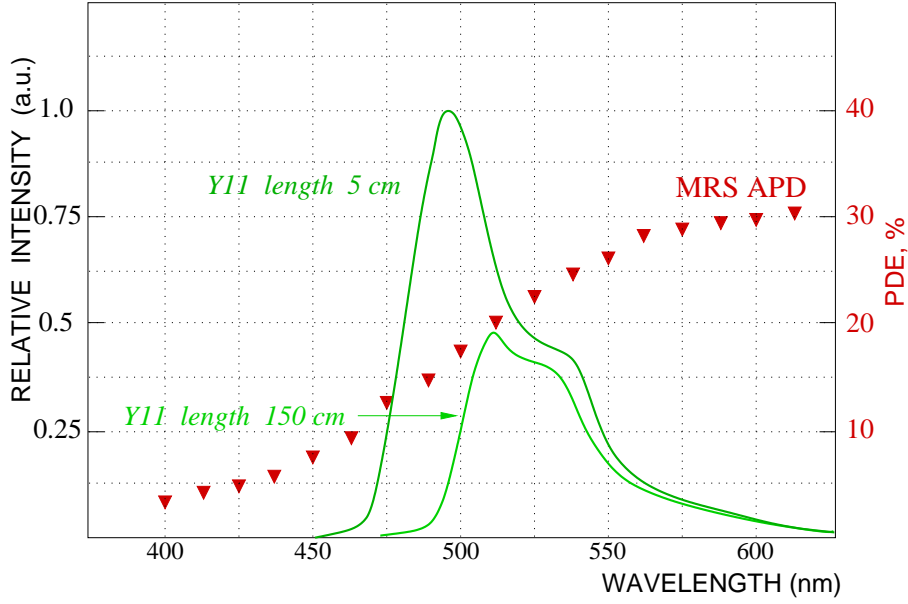}
\caption{Light emission spectrum of the Y11(150) fiber (from Kuraray 
plots~\cite{kuraray}) 
and the PDE of a  MRS photodiode.}
\label{fig:y11_sipm} 
\end{figure}

The PDE dependence on the wavelength of the detected light, as well as the 
emission spectrum of the WLS fiber Y11  are 
shown in Fig.~\ref{fig:y11_sipm}. 
 The peak emission of the fiber depends on its length due to the absorption of 
 the light. The spectral response of a MRS photodiode  was measured 
in a spectrophotometer  calibrated with a PIN--diode~\cite{si3}.
The PDE was measured at higher  $V_{bias}$ and, therefore,   
the dark rate was higher (about 2.3 MHz  for a discriminator 
threshold of  0.5 p.e.).   
The PDE decreases by about 50\%  when  $V_{bias}$  is lowered such that 
the dark rate is   $\sim 1$ MHz.

\begin{figure}[htb]
\centering\includegraphics[width=12cm,angle=0]{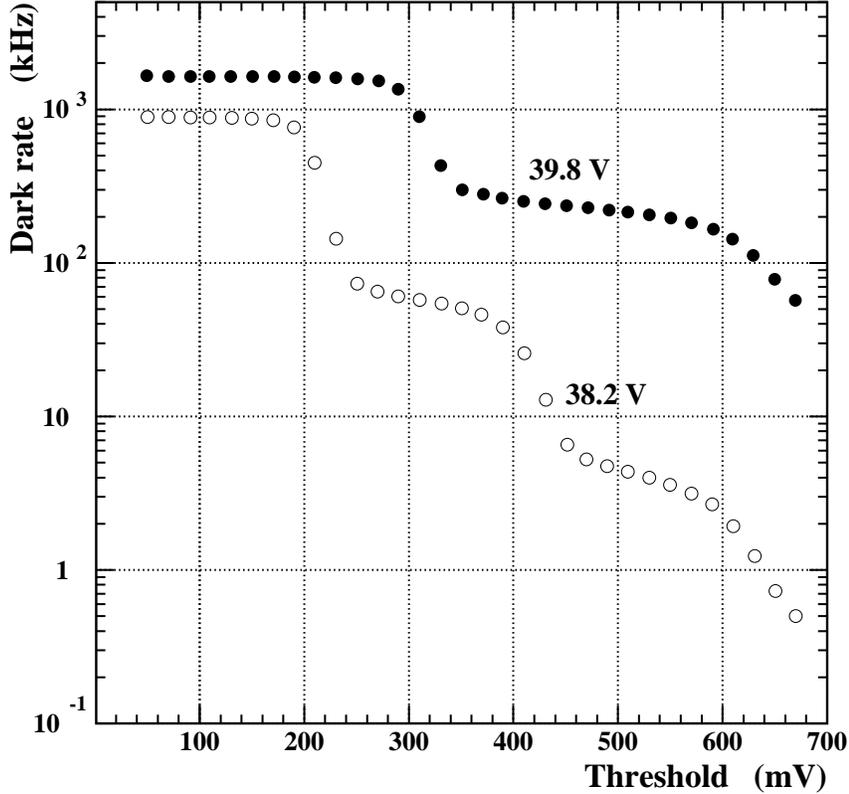}
\caption{Dark  rate vs the discriminator threshold for two bias voltages. 
The discriminator threshold values of 100, 300, and 500 mV correspond to 
0.5 p.e., 1.5 p.e., and 2.5 p.e., respectively, for $V_{bias} = 38.2$~V. }
\label{fig:dark-rate} 
\end{figure}

{\it Dark rate.}
The limiting factor for a wide application of   the MRS photodiodes in 
the readout of scintillators is the  dark noise rate 
which originates from 
  thermally created carriers in the depletion region under high electric
fields. The dark rate mainly consists of  single p.e. pulses. Larger amplitude 
pulses also contribute
 to the dark rate, as shown in Fig.~\ref{fig:dark-rate}. 
However,  
the intensity of the  pulses with 2 p.e. amplitudes is about 15 times   less 
than the intensity 
of the single p.e. pulses. Large amplitudes ($> 1$ p.e.) are generated by 
the optical 
cross-talk as well as by accidental pile-ups between the independent pixels, 
though the second 
effect is relatively small. The dark rate decreases  to the level of a few kHz 
for the threshold of 2.5 p.e.   The intensity of 
1 p.e. pulses  as well as 2 p.e. pulses is significantly higher 
for higher bias voltage, as can be seen from Fig.~\ref{fig:dark-rate} in case 
of $V_{bias} = 39.8$~V. 
Dark pulses and mostly leakages create  the 
dark current through a MRS photodiode in
 a typical range of $0.3 - 1.0~\mu A$. 
The dark rate decreases to about 1~kHz/mm$^2$ at $-70^{\circ}$C, 
i.e. decreases by a factor of 
 2  for every temperature drop of  $9 - 10^{\circ}$C   for a threshold of 0.5 p.e. and 
and the  condition of keeping the gain   at a constant value by tuning 
the bias voltage.

{\it Optical cross--talk.}
A single incident photon can create more than a single p.e. due to 
optical cross--talk. The mechanism of this effect is not quite clear. 
Models predict that during the Geiger discharge and recombination of carriers 
infrared photons are emitted. These photons penetrate into the  adjacent pixels 
and fire them. Optical cross--talk leads to a higher than expected signal 
as Geiger discharge occur in additional pixels. 
The absolute value of the cross--talk can be calculated by assuming a  Poisson 
distribution for the number of photoelectrons observed in response 
to the light from a  LED. When the MRS 
photodiode noise is 
small, the measured mean number of fired pixels  by the LED photons, 
${\bar N}_{LED}$, 
is  compared with the calculated value ${\bar N}_{pe}$   given by
\begin{equation}
{\bar N}_{pe} = - lnP(0), 
\label{eq:poisson}
\end{equation} 
where $P(0)$ is the fraction of 0 p.e., or ``pedestal'' events. Then
the deviation of the ratio ${\bar N}_{LED}/{\bar N}_{pe}$ from 1 
 gives the cross--talk value. These values are 
 presented for several bias voltages   in Fig.~\ref{fig:pde}.
A cross--talk value of  about 5\%  was 
obtained at a  bias voltage that provides a dark noise rate of 
$\leq 1$ MHz 
at a discriminator threshold of 0.5 p.e.  The cross--talk  is larger for 
higher bias voltage. The sharp decrease of the dark rate shown in 
Fig.~\ref{fig:dark-rate} is a good 
demonstration of the low optical  cross--talk. It should be noted that 
the signal amplitude  in photoelectrons is less than  the measured 
amplitude determined 
as  the average number of fired pixels by the cross-talk value, 
i.e. by a few per cent.
 
\begin{figure}[htb]
\centering\includegraphics[width=12cm,angle=0]{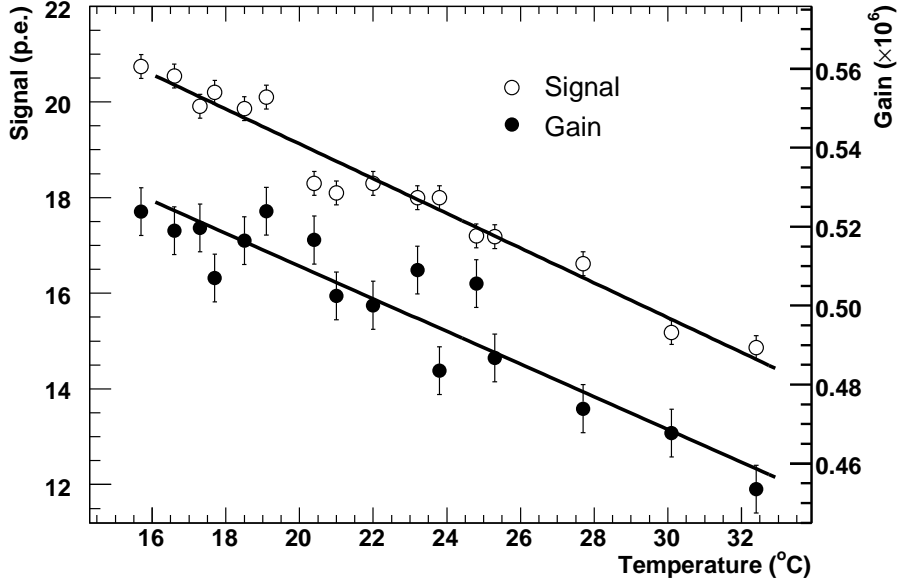}
\caption{The MRS  signal from a green LED  and gain as a function of   
the ambient air temperature.}
\label{fig:signal_gain_temp} 
\end{figure}

\begin{figure}[htb]
\centering\includegraphics[width=12cm,angle=0]{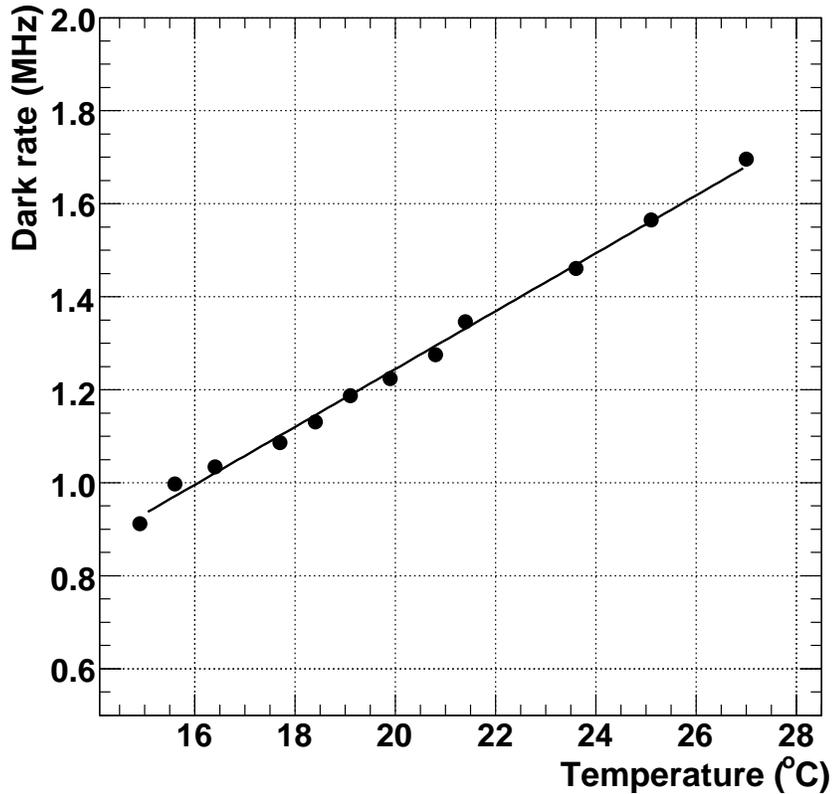}
\caption{The  dark rate for threshold of 0.5 p.e. vs 
the  ambient air temperature.}
\label{fig:dark_temp} 
\end{figure}

{\it Temperature dependence.}
The PDE and gain of MRS photodiodes (as well as the  signal amplitude) 
are expected to be
sensitive to temperature because the breakdown voltage   depends on
temperature. The MRS signal amplitude (light yield) is proportional to 
$N_{\rm photons}\times 
{\rm PDE}\times {\rm gain}$, where  $N_{\rm photons}$ is the number of photons 
from the  LED which illuminate the photodiode.
The parameters of the MRS photodiodes  were measured over temperatures 
from 15$^{\circ}$ to 33$^{\circ}$C.  The temperature variation of 
the  MRS signal (a green LED was used as a light source)
is presented in Fig.~\ref{fig:signal_gain_temp}.
 The  MRS signal   dependency of -1.5~\%/$^{\circ}$C is obtained for 
 increasing temperature.
 The MRS gain itself  decreases with temperature as   -1.2~\%/$^{\circ}$C 
 (see Fig.~\ref{fig:signal_gain_temp}), while the PDE varies with temperature 
 as -0.3~\%/$^{\circ}$C.
 The  dark rate  depends on temperature with a coefficient  of 
 62 kHz/$^{\circ}$C, as shown in Fig.~\ref{fig:dark_temp}.
  Decreasing the temperature
below $0^{\circ}$C greatly reduces the noise and increases the PDE, 
as shown in Ref.~\cite{si3}.

\begin{figure}[htb]
\centering\includegraphics[width=11cm,angle=0]{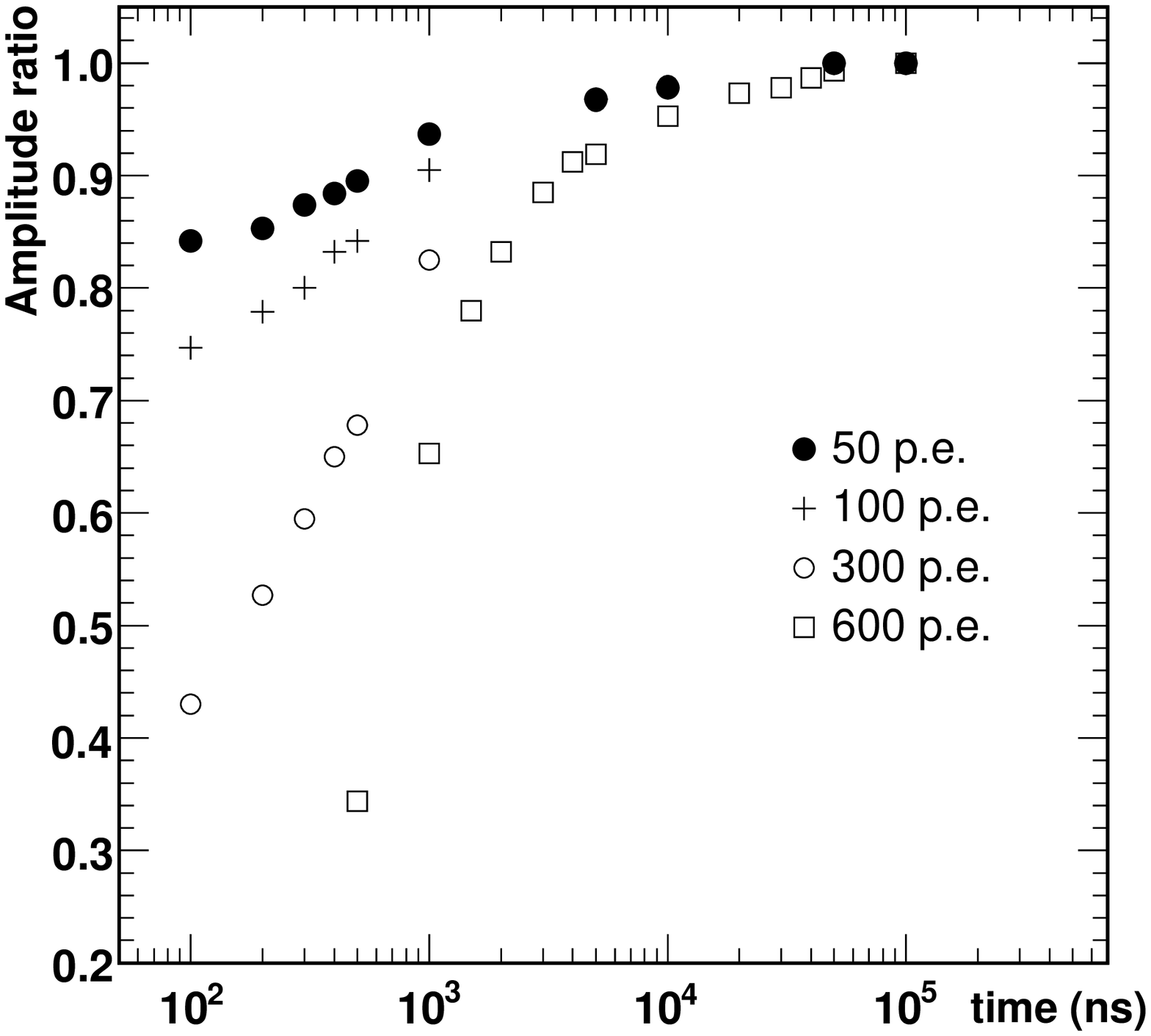}
\caption{The relative amplitude of the MRS delayed signal from a LED 
 as a function 
of the time difference between the first LED pulse and the second one. 
The   amplitude is 
normalized to its value when the first LED signal is off.   The
amplitudes of 
the first  signal are  50, 100, 300, 600 photoelectrons.}
\label{fig:recovery} 
\end{figure}

{\it Recovery time.}
The ability of MRS photodiodes to operate at high counting rates was tested 
using 
two LED signals. We measured the amplitude of the second signal $A_2(t)$ as a 
function of 
the time difference $t$ between the first and second  signal. 
Fig.~\ref{fig:recovery} shows the  ratio 
$A_2(t)/A_2(0)$, where $A_2(0)$ is the amplitude of the second signal when 
the first LED
signal is off.  
As seen from Figure 9, a minimum time interval between the first 
and second pulse of about 5 $\mu$s is needed 
 to recover about 95\% of the full amplitude 
of the second signal. This long recovery time is due to the fact that 
the individual 
resistor of each pixel, $R_{pixel}$, has the value of about 20 M$\Omega$, 
the pixel capacitance 
$C_{pixel}$ is 
typically 50 fF that gives $\tau = R_{pixel}\cdot C_{pixel} \sim 1 \mu$s. 
 
\begin{figure}[htb]
\centering\includegraphics[width=12cm,angle=0]{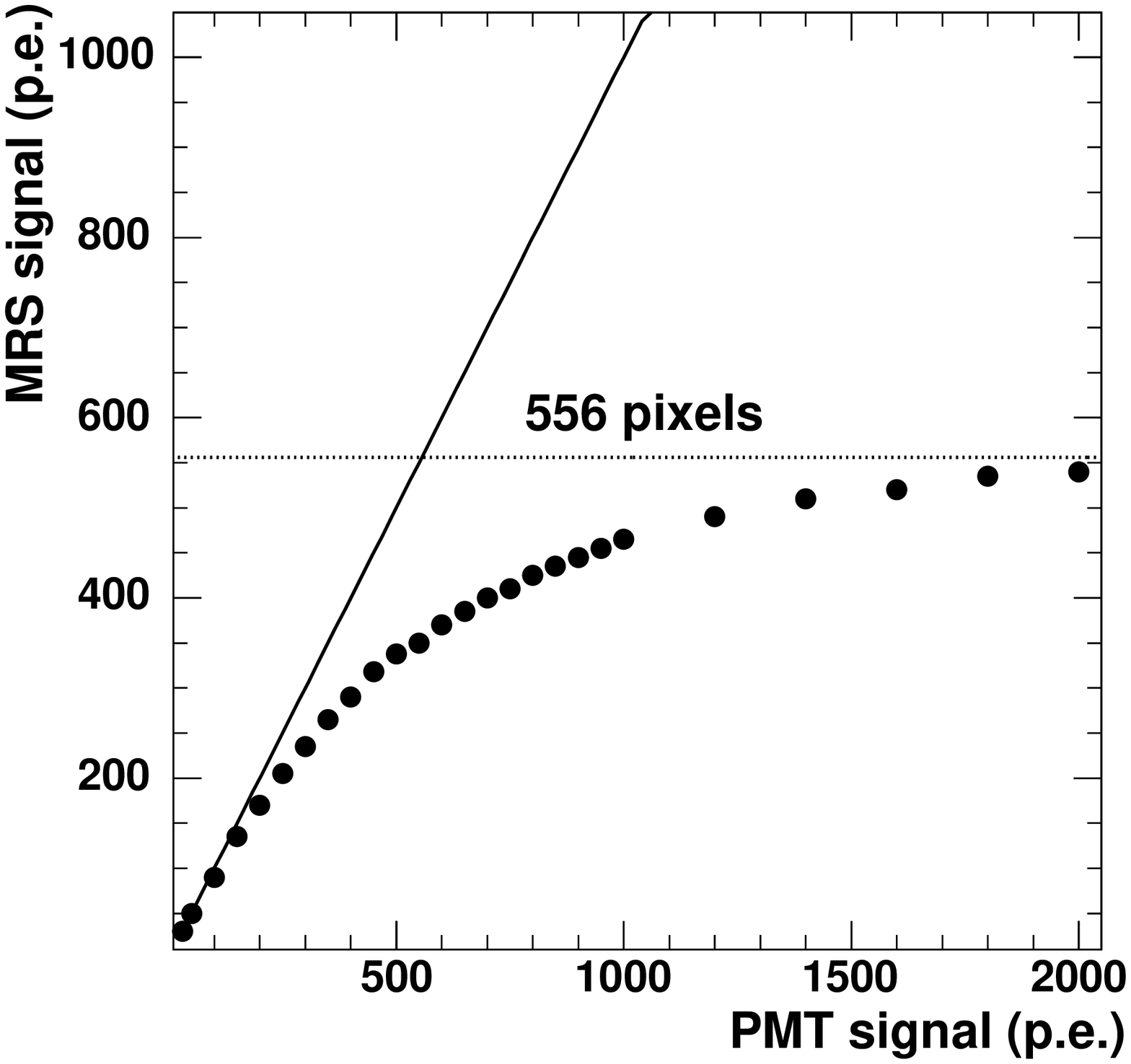}
\caption{The MRS photodiode response vs the PMT signal. The signal of the 
MRS photodiode is saturated at the level of the maximum number of 
 556 pixels.}
\label{fig:nonlin} 
\end{figure}

{\it Dynamic range and linearity.} 
The  dynamic range of the MRS photodiode is limited by the finite number of 
pixels. The saturation of the  MRS photodiode in response to large light signals 
is shown in Fig.~\ref{fig:nonlin}.  
In this test, the LED signal was ajusted to obtain similar 
response for both PMT and MRS photodiode at the light yield below 150 p.e. 
The photodiode
signal is linear for low amplitutes.  A nonlinearity of about 10\% is already
seen for a signal of 150--200~p.e. 
For the PMT signal of 500~p.e., the MRS photodiode has  
 a 30\%  smaller signal than that of the PMT. Full saturation of the 
 556 pixel MRS photodiode starts at a reference (PMT) signal of more 
 than  2000~p.e.

\begin{figure}[htb]
\centering\includegraphics[width=10cm,angle=0]{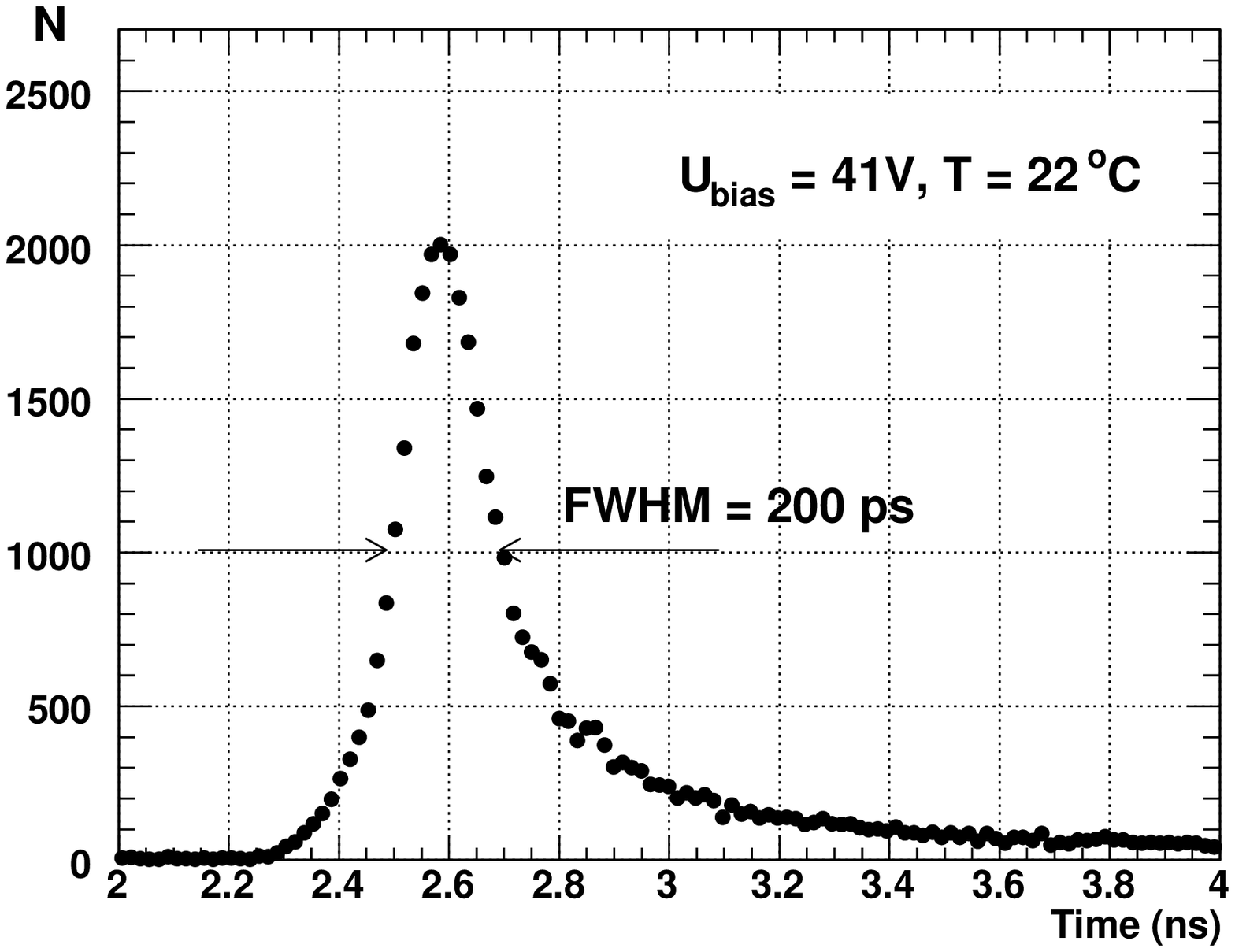}
\caption{Time spectrum of  single photoelectron signals obtained by a MRS
photodiode at $22^{\circ}$C.}
\label{fig:timing_mrs} 
\end{figure}

{\it Timing.}
The development of the Geiger discharge in a small depth ($\sim 0.8~\mu m$) 
of the depletion region takes a few hundred picoseconds. The typical rise time 
is 1~ns, the decay time is
determined by the pixel capacitance.   A laser with a wavelength of 635~nm and 
a pulse width of 35~ps (fwhm) was used to measure  the intrinsic  time 
resolution of  1 p.e. pulses. Very weak laser light created only 
1 p.e. signals in a MRS photodiode. The time resolution obtained for a 
threshold of 0.2 p.e.  at $22^{\circ}$C is presented in 
Fig.~\ref{fig:timing_mrs}. 
 
\begin{figure}[htb] 
\centering\includegraphics[width=15cm,angle=0]{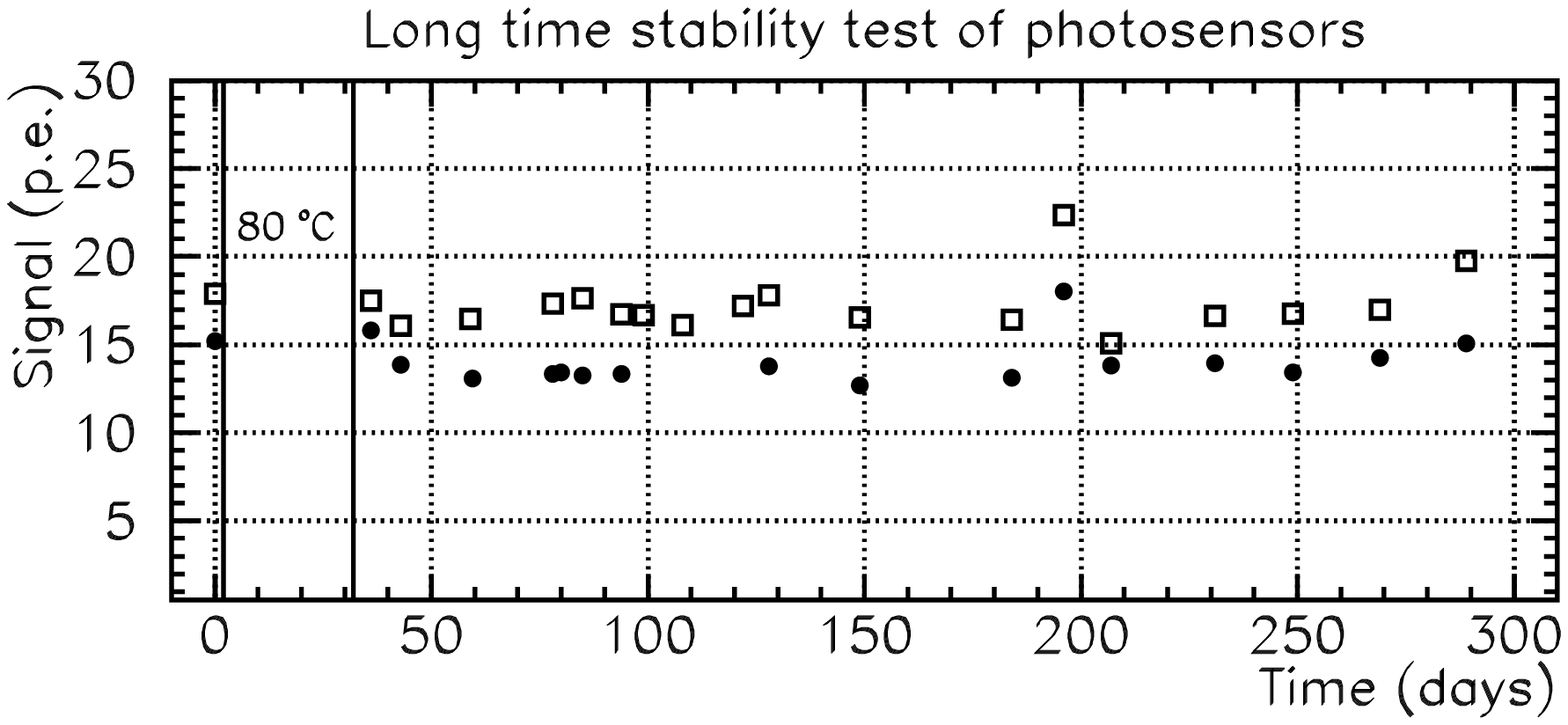}
\caption{Light yields  of two MRS photodiodes measured 
with a green LED before and after the heating test. The heat exposure 
at 80$^{\circ}$C lasted for 30 days. Successively, the photodiodes were 
monitored for 260 days. The signals were corrected for temperature 
changes in the range from $18 - 27^{\circ}$C. Each point has an error 
of $\pm 2.0$ p.e.}
\label{fig:oven} 
\end{figure}

{\it Life time.} 
The failure rate of the MRS photodiodes is an important figure of merit for 
the overall detector performance, because most photodiodes can not be 
replaced without significant 
disassembly of  the  ND280 detector.
Exposures to elevated temperature  are used to evaluate the expected life time of 
semiconductor devices. We have placed 19 MRS photodiodes  in an oven 
at 80$^{\circ}$C for 30 days. All photodiodes were kept under bias 
voltages to provide 
the same dark current as at  room temperature. One of the devices 
started to conduct a large current after a week of heating. 
The failed device had the worst PDE among of the tested 
devices. This points to a possible defect in its structure.  
All other devices   
passed the test without residual effects, and the LED signals  measured 
by these MRS diodes after a period of 30 days of elevated  
temperatures did not show any significant 
degradation. 
The  signals of the tested photodiodes  in response to a LED photons 
were measured before  the heating and  
for a period of about 260 days after the heating. The results for two 
devices  are presented  in  Fig.~\ref{fig:oven}. 
The accuracy of the l.y. measurements is about 2.0 p.e.~(rms) due to 
misalignments between the fiber and the MRS photodiodes that are estimated to 
be between 20 and 100 $\mu$m in this series of tests.  
For more than  7 months after  heating no degradation in the light yield was observed.

Two hundred MRS photodiodes fabricated from a few different wafers were tested.
In order  to meet 
the requirements of the experiment, the dark rate of each device 
should be kept  close to
a reference 
value of   1 MHz at room temperature 
22$^{\circ}$C and  for a  discriminator threshold of 0.5 p.e. 
This required the individual adjustment of the bias voltage of 38 
to 42.1~V for each 
photodiode, that in turn  resulted in  a wide 
range  of 8.9--18.8\% for the PDE values,  and the  
gains of these 200 MRS photodiodes were found to be between    
$0.34\times 10^6$ and  $0.69\times 10^6$.

\section{Scintillator detectors with WLS fiber readout}

\begin{figure}[htb] 
\centering\includegraphics[width=11cm,angle=0]{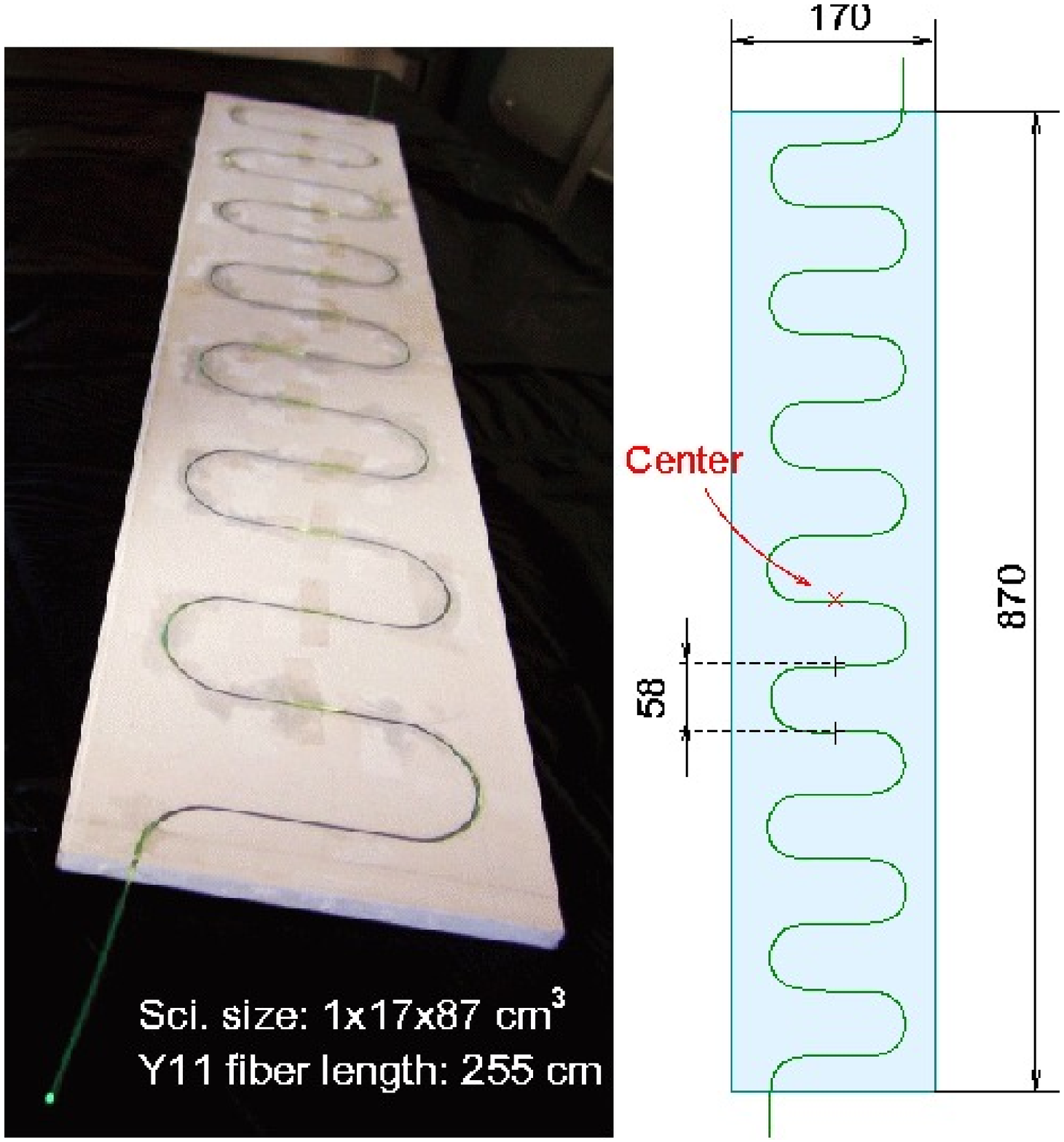}
\caption{Scintillator slab with S--shaped fiber readout: photograph (left) and 
schematic view (right).}
\label{fig:s-counter} 
\end{figure}

Several subdetectors of  the T2K near detector complex will be composed 
 of many 
scintillator detectors: 
rectangular and triangular scintillator bars, and various scintillating 
slabs. 
All these elements use  embedded WLS fibers to read light from the 
scintillators. 
A well known design will be adopted  for  the  bars with WLS fibers: 
one straight hole or groove on one surface for a WLS fiber.  For large 
scintillator
slabs (SMRD,  active elements of Ecal), the usage of the standard readout 
scheme with 
several
equidistant WLS fibers which run along the slab is not appropriate due to 
mechanical constraints  of the UA1 magnet. 
We consider   
extruded scintillator slabs with a double--ended WLS fiber readout technique as 
active elements for the SMRD detector. Instead of a few parallel  WLS fibers 
 we propose to use a single long WLS fiber embedded in an 
S--shape groove 
which reduces the maximum path length that light has to travel within 
the scintillator to a few  cm.  
 The detector prototypes   
were manufactured using an extrusion technique 
developed at the Uniplast Factory, Vladimir, Russia. The scintillator is 
etched by a chemical agent that results in the formation of a micropore 
deposit over the plastic surface. The thickness of the 
deposit (30--100 $\mu$m) depends on the etching 
time.   Details can be found in 
Ref.~\cite{extrusion}. 
  A scintillator slab of 1x17x87~cm$^3$  with an 
S--shaped groove  of 3~mm depth was  manufactured for a test in a pion/proton
beam. 
The half--period of the S--shape groove   
 is 58~mm, as shown in Fig.~\ref{fig:s-counter}. 
A 2.55~m long multi--clad  Kuraray Y11 (200 ppm dopant) WLS fiber of 1~mm 
diameter is 
embedded into the groove with an optical grease and both ends are attached to 
MRS photodiodes. To avoid the 
 degradation of the Y11 parameters in real SMRD counters, the bending procedure to make 
 3 cm radius loops will be done at 
 the temperature of about 80$^{\circ}$C.    
 
This module was first tested with cosmic muons. 
A light yield (l.y.) of 16.4~p.e. was obtained for MIP's 
in the center for  summed signals from both photodiodes. In order 
to suppress the timing spread caused by the trigger counters the 
combination  $(TDC_{left} - TDC_{right})/2$ was used to measure the time
resolution.  A time resolution of $\sigma = 1.57 $ ns  was obtained 
for MIP's which passed through the central part of the slab.

\section{Beam tests of the SMRD  prototype}
\subsection{Beam test setup}
\begin{figure}[htb] 
\centering\includegraphics[width=13cm,angle=0]{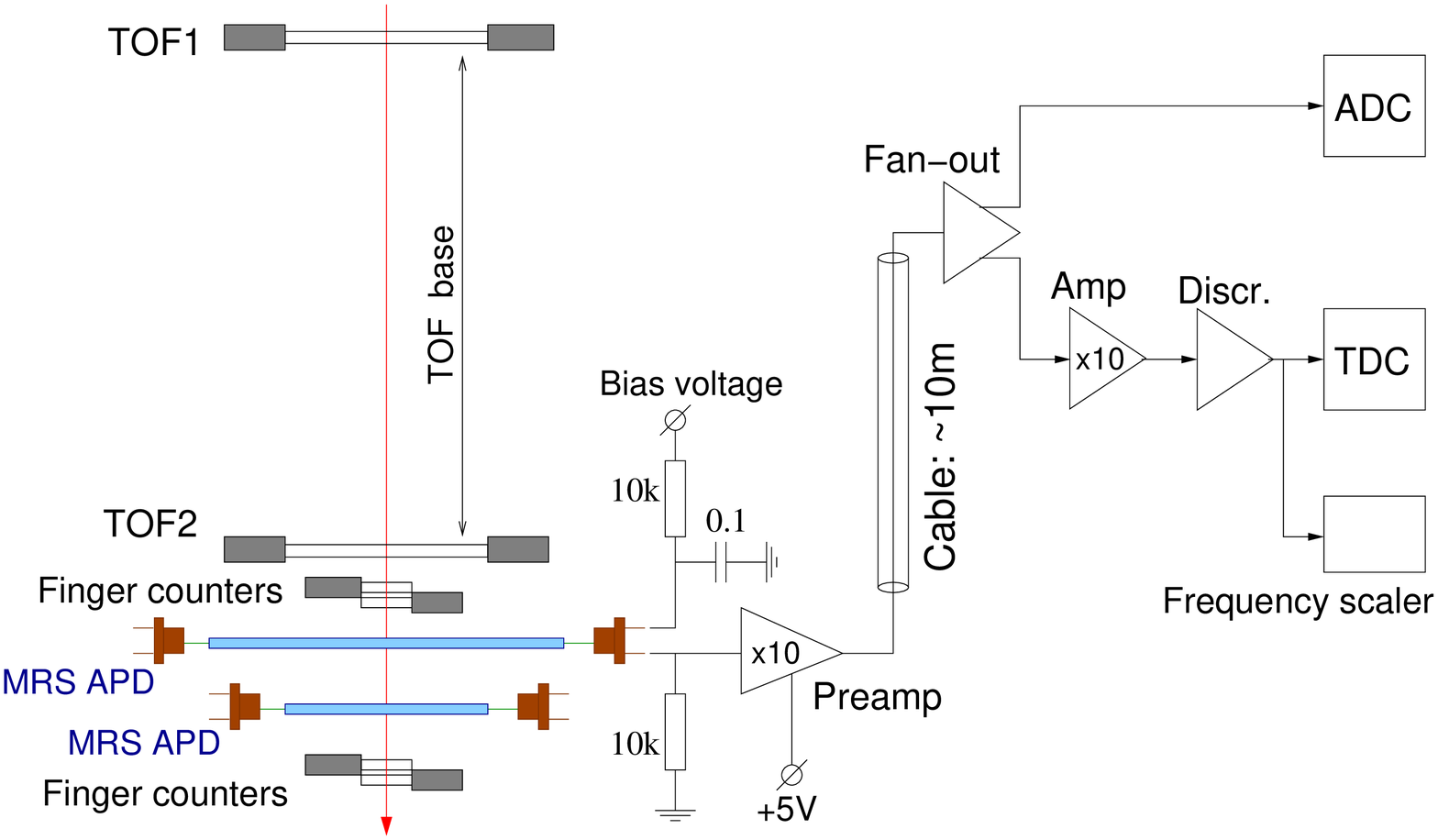}
\caption{The electronic readout diagram used in the beam test.}
\label{fig:electronics} 
\end{figure}

A beam test of extruded scintillators with embedded WLS fibers was performed 
at the KEK 12-GeV proton synchrotron 
with 1.4 GeV/c protons and pions. Two counters were placed in a beam and 
tested simultaneously. One was the S--grooved scintillator (S--counter) 
described in the previous section, another one was a scintillator 
slab,  1x18x50~cm$^3$ in size, with a single straight groove 
in the middle of the plastic (W--counter). An 1.11~m long Kuraray Y11(200) fiber 
of 1~mm 
diameter was
embedded with  optical grease  into the straight groove  of the 
W--counter. The fiber 
in each counter is viewed from  
both ends by  MRS photodiodes. The fibers  are 
directly coupled  to  photosensors  
inside special sockets. The  detectors were mounted on a
platform which could be moved horizontally and vertically with 
respect to the beam line. Upstream of the counters, a TOF system was used 
to separate pions and protons. Finger trigger counters restricted 
the beam spot size  to a  $10\times 10 {\rm mm}^2$ square.   The signals 
were amplified by 
fast hybrid preamps mounted 
directly  behind the photodiodes. The bias voltage was set individually for 
each MRS photodiode to limit the dark rate to about 1.2~MHz for a
0.5 p.e. threshold. A simplified electronic diagram of the beam test setup is shown in 
Fig.~\ref{fig:electronics}. 

\begin{figure}[htb] 
\centering\includegraphics[width=12cm,angle=0]{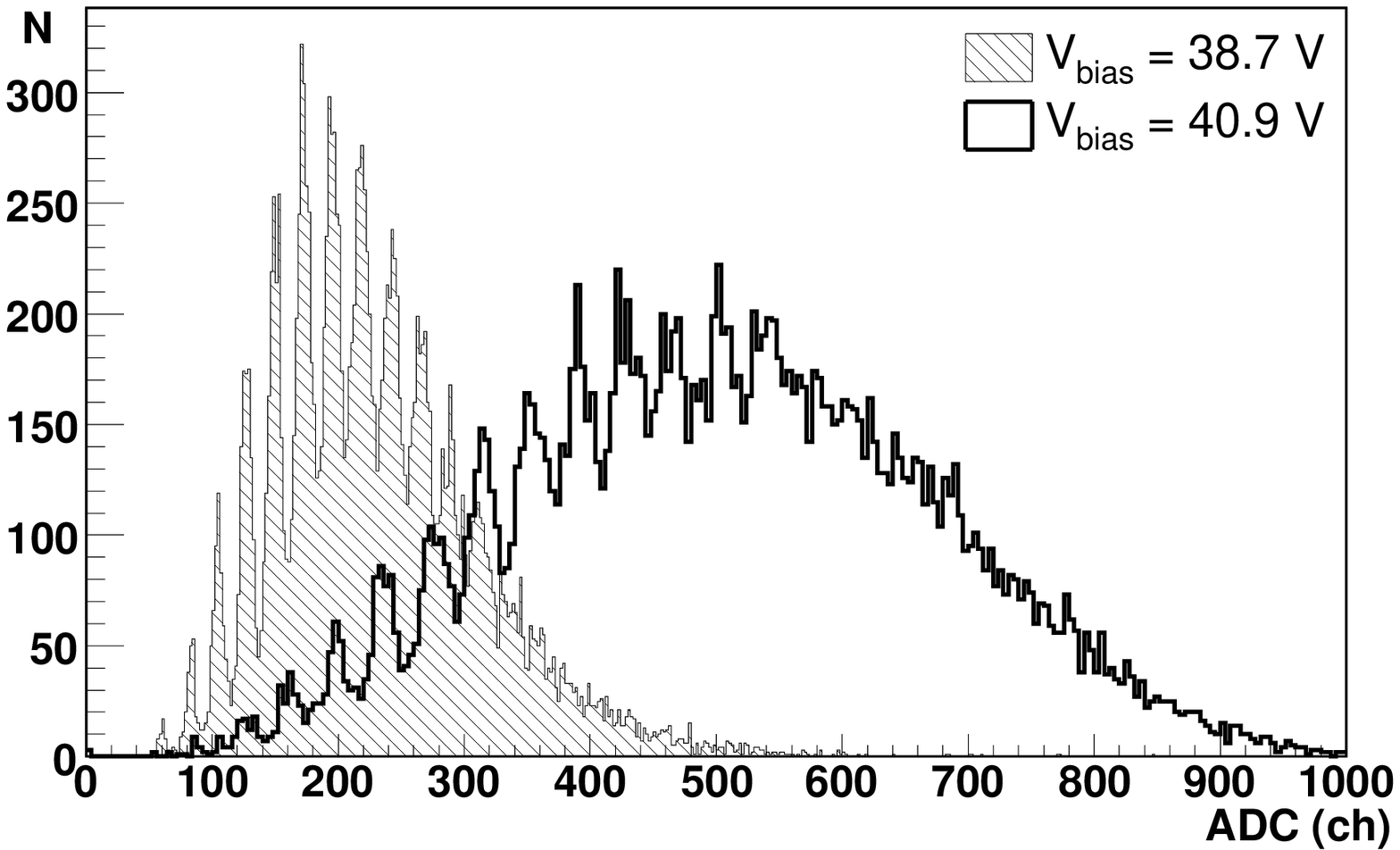}
\caption{The spectra of 1.4 GeV/c pions measured at two bias voltages.}
\label{fig:spectra-vs-bias} 
\end{figure}

Only pions were selected for analysis as their response is close to that 
expected from MIP's in the T2K experiment. The ambient temperature 
 drifted between 15--18$^{\circ}$C during the beam 
test. Typical ADC spectra from pions are  
shown in Fig.~\ref{fig:spectra-vs-bias}. 
In this 
Figure, one spectrum is obtained with a bias voltage of 
38.7~V which corresponds to a  dark rate of 0.86~MHz for a
discriminator threshold of 0.5 p.e. The 
second spectrum is measured with a bias voltage of 40.9~V corresponding to
a dark rate of almost 2~MHz.
The individual photoelectron peaks are  clearly separated  at 38.7~V 
and still visible at 40.9~V. 

\subsection{Detection efficiency of MIP's}
The light yield scan over the S--counter surface is presented in 
Table~\ref{table:s-ly}.  
\begin{table}[htbp]
\caption{Light yield (p.e./MIP)  
over the S--counter.  Sum of the signals from  both ends.
The bias voltage of 38.7~V was applied to both MRS photodiodes.}
\vspace{.5cm}
\begin{center}
\begin{tabular}{|c|c|c|c|c|c|c|c|c|c|}
\hline
 ~~~$y$, mm & -80 & -60 & -40 & -20 & 0 & 20 & 40 & 60 & 80  \\
$x$, mm~~~  &\       &\      &\       &\     &\    &\  &\   &\   &\   \\
\hline
-406    & 6.3 & 11.9 & 13.2 & 16.0 & 17.8 & 18.6 & 18.7 & 18.8 & 13.8  \\
\hline
-319   & 10.2 & 11.6 & 14.7 & 15.7 & 15.9 & 16.4 & 17.3 & 16.7 & 13.2   \\
\hline
-200   & 10.2 & 12.7 & 14.3 & 16.0 & 15.5 & 16.6 & 18.2 & 18.7 & 15.0   \\
\hline
-87    & 10.4 & 11.7 & 13.0 & 15.0 & 15.0 & 15.4 & 16.3 & 14.5 & 11.0   \\
\hline
-30     & 8.8 & 14.4 & 16.4 & 16.2 & 15.3 & 14.6 & 13.8 & 12.9 & 10.9   \\
\hline
0      & 11.0 & 12.9 & 14.8 & 14.3 & 15.3 & 16.2 & 14.9 & 13.9 & 11.7   \\ 
\hline
30      & 9.3 & 11.8 & 12.6 & 14.4 & 14.8 & 15.8 & 16.8 & 16.0 & 13.0   \\
\hline
87     & 12.3 & 14.4 & 15.1 & 14.6 & 14.2 & 14.7 & 14.2 & 12.9 & 11.8   \\
\hline
200    & 11.0 & 15.2 & 16.0 & 15.9 & 15.6 & 15.5 & 13.7 & 12.9 & 11.3   \\
\hline
319    & 12.6 & 16.1 & 17.4 & 16.3 & 15.5 & 15.2 & 14.5 & 12.3 & 10.4  \\
\hline
406    & 11.8 & 12.9 & 17.9 & 19.9 & 20.0 & 19.1 & 19.1 & 16.2 & 11.9  \\
\hline
\end{tabular} 
\end{center}
\label{table:s-ly}  
\end{table}
The beam spot spreads beyond the scintillator area near the edges 
at $y = \pm 80$ mm 
causing the small l.y. values.  If the edges are ignored,  the light 
output over the 
S--counter (sum of both end signals) varies from 12 to 20~p.e./MIP. The largest l.y. 
is measured at the 
ends, close to either  of the two MRS photodiodes. In order to obtain 
the detection 
efficiency the ADC 
spectra were analyzed. The event is considered accepted  if its 
ADC amplitude exceeds a certain threshold set in number of p.e.  The 
average statistics 
in each location is about 2000 events.
Table~\ref{table:s-3}
\begin{table}[htbp]
\caption{Detection efficiency  over the S--counter for a threshold of 2.5 p.e.  
The bias voltage of 38.7~V was applied to both MRS photodiodes.}
\vspace{.5cm}
\begin{center}
\begin{tabular}{|c|c|c|c|c|c|c|c|c|c|}
\hline
 ~~~$y$, mm & -80 & -60 & -40 & -20 & 0 & 20 & 40 & 60 & 80  \\
$x$, mm~~~  &\       &\      &\       &\     &\    &\  &\   &\   &\   \\
\hline
-406    & 0.660 & 0.995 & 0.999 & 1.000 & 1.000 & 0.999 & 1.000 & 1.000 & 0.999 \\
\hline
-319   & 0.750 & 0.999 & 0.999 & 0.999 & 0.999 & 0.999 & 0.999 & 0.998 & 1.000   \\
\hline
-200   & 0.788 & 0.998 & 0.999 & 0.999 & 1.000 & 1.000 & 0.999 & 0.999 & 1.000   \\
\hline
-87    & 0.839 & 0.996 & 0.998 & 0.999 & 1.000 & 0.999 & 1.000 & 0.999 & 0.995  \\
\hline
-30     & 0.886 & 0.999 & 0.999 & 1.000 & 1.000 & 0.999 & 1.000 & 0.997 & 0.998   \\
\hline
0      & 0.985 & 0.998 & 0.997 & 0.998 & 0.996 & 0.998 & 0.999 & 0.999 & 0.998   \\ 
\hline
30      & 0.918 & 0.996 & 0.998 & 0.999 & 1.000 & 1.000 & 1.000 & 0.999 & 0.998  \\
\hline
87     & 0.989 & 0.997 & 1.000 & 0.999 & 0.999 & 1.000 & 0.998 & 0.999 & 0.996   \\
\hline
200    & 0.995 & 0.998 & 1.000 & 0.999 & 1.000 & 1.000 & 1.000 & 0.999 & 0.994   \\
\hline
319    & 0.999 & 0.998 & 1.000 & 1.000 & 1.000 & 1.000 & 1.000 & 0.998 & 0.994  \\
\hline
406    & 0.998 & 0.999 & 1.000 & 1.000 & 1.000 & 0.999 & 1.000 & 1.000 & 0.984  \\
\hline
\end{tabular} 
\end{center}
\label{table:s-3} 
\end{table}
demonstrates the S--counter pion detection efficiency 
 when the threshold for the sum of both end signals is set to 2.5~p.e. 
   The detection efficiency is 
 close to 100~\% except 
 for the edge area where a part of the beam missed the counter due to  
 some misalignment between the beam counters and the tested detector. 
 For a higher threshold of 4.5~p.e set for the sum of the amplitudes from 
 the two ends, the MIP detection efficiency is greater than 98\%.  
  We can  conclude that  the l.y. of more 
 than 12~p.e. (sum of both ends)  satisfies the requirement  for the  
 S--counter
 to provide a detection 
 efficiency greater than 99\% for a MIP.  
 If we require that each MRS photodiode signal exceeds   0.5~p.e. 
  the MIP detection  efficiency is found  to  
 be about 99.5\%.

\begin{figure}[htbp]
\centering\includegraphics[width=14cm,angle=0]{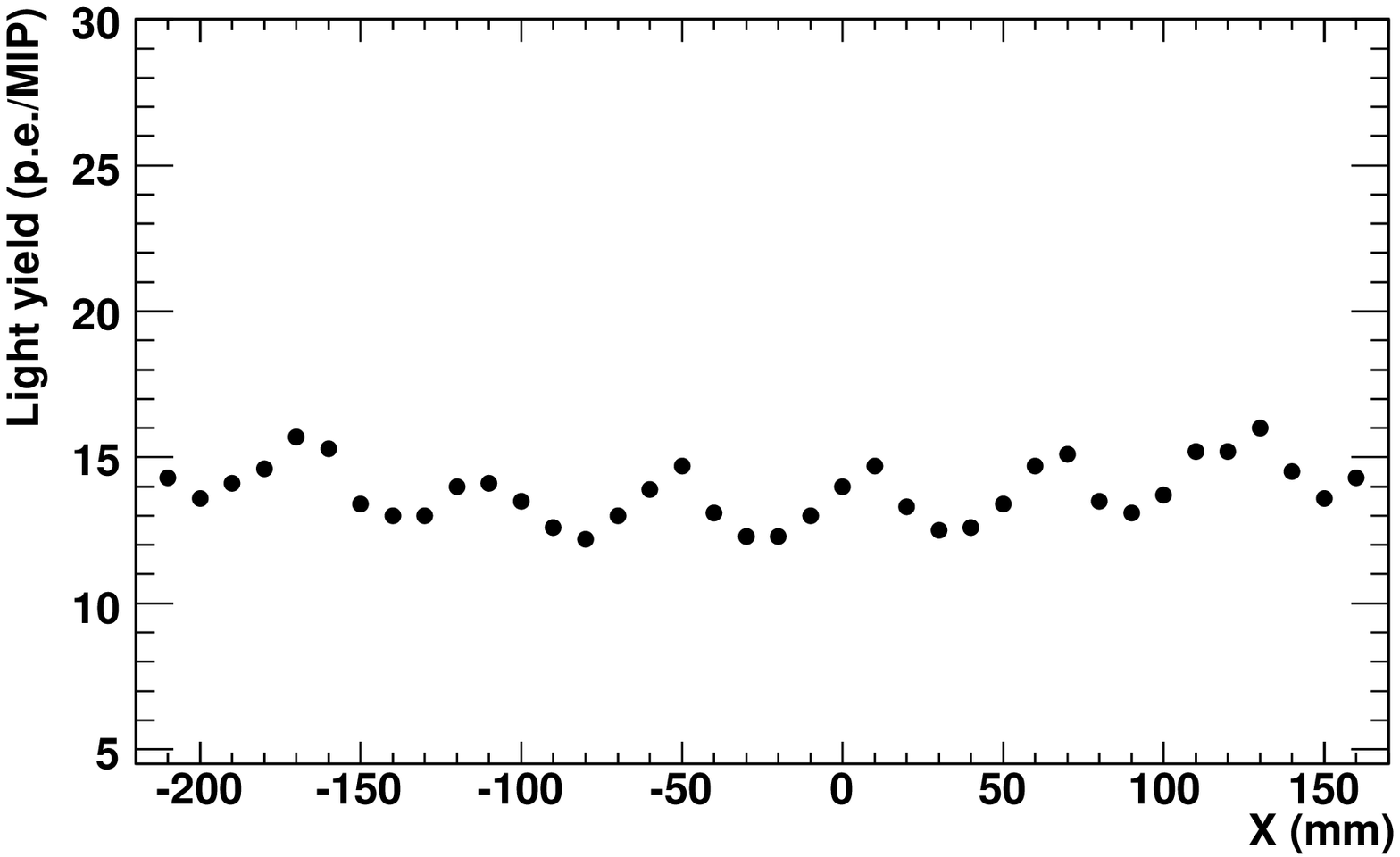}
\caption{The light yield scan of the S--counter along the x-axis. The 
beam spot 
size 
is $0.5\times 0.5$~cm$^2$, the scan step size is 1~cm. The sine--like waveform 
corresponds to the 
WLS fiber route with 58~mm spacing between neighboring segments.}
\label{fig:scan-x-S} 
\end{figure}

\begin{figure}[htbp] 
\centering\includegraphics[width=14cm,angle=0]{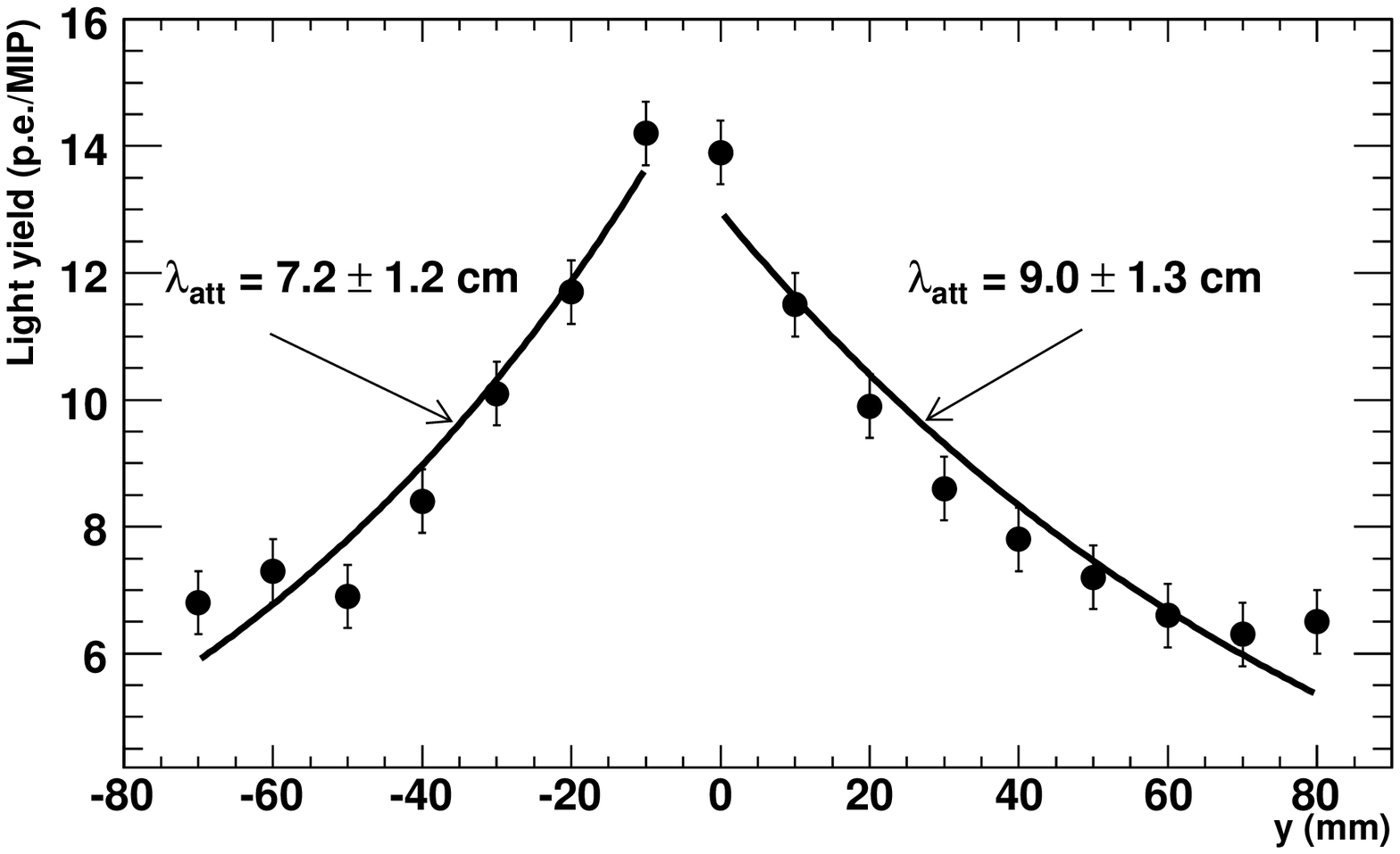}
\caption{The light yield scan of the W-counter along the  y--axis , that is 
 perpendicular to the 
fiber direction. The beam spot size is $0.5\times 0.5$~cm$^2$, and the scan step 
size is 1~cm.}
\label{fig:scan-y-W} 
\end{figure}

To make a detailed scan along the middle line of the S--counter the 
size of the beam spot  was reduced to 
$0.5\times 0.5$~cm$^2$.  The result is shown in 
Fig.~\ref{fig:scan-x-S}. 
The WLS fiber route is clearly reflected in the pattern of the l.y. distribution 
along the x--axis. A maximum l.y. is observed in points where the beam crosses 
the fiber, while   the l.y. drops by about 20\% in between 
the fiber segments. Fig.~\ref{fig:scan-y-W} 
shows the result of the  transverse l.y. scan across the W--counter, 
which has a single 
straight groove running down the middle of the counter. The attenuation length of 
scintillation light  before capture by the WLS fiber is obtained to be 
$9.0 \pm 1.3$ cm and $7.2 \pm 1.2$ cm for the upper and the lower part of the
W--counter, respectively. These values are consistent within measurement
uncertainty, and 
the average attenuation length of such a counter is estimated to be about 
$8.1\pm 0.9$ cm.

\subsection{Time and spatial resolution of the S--counter}
\begin{figure}[htbp] 
\centering\includegraphics[width=10cm,angle=0]{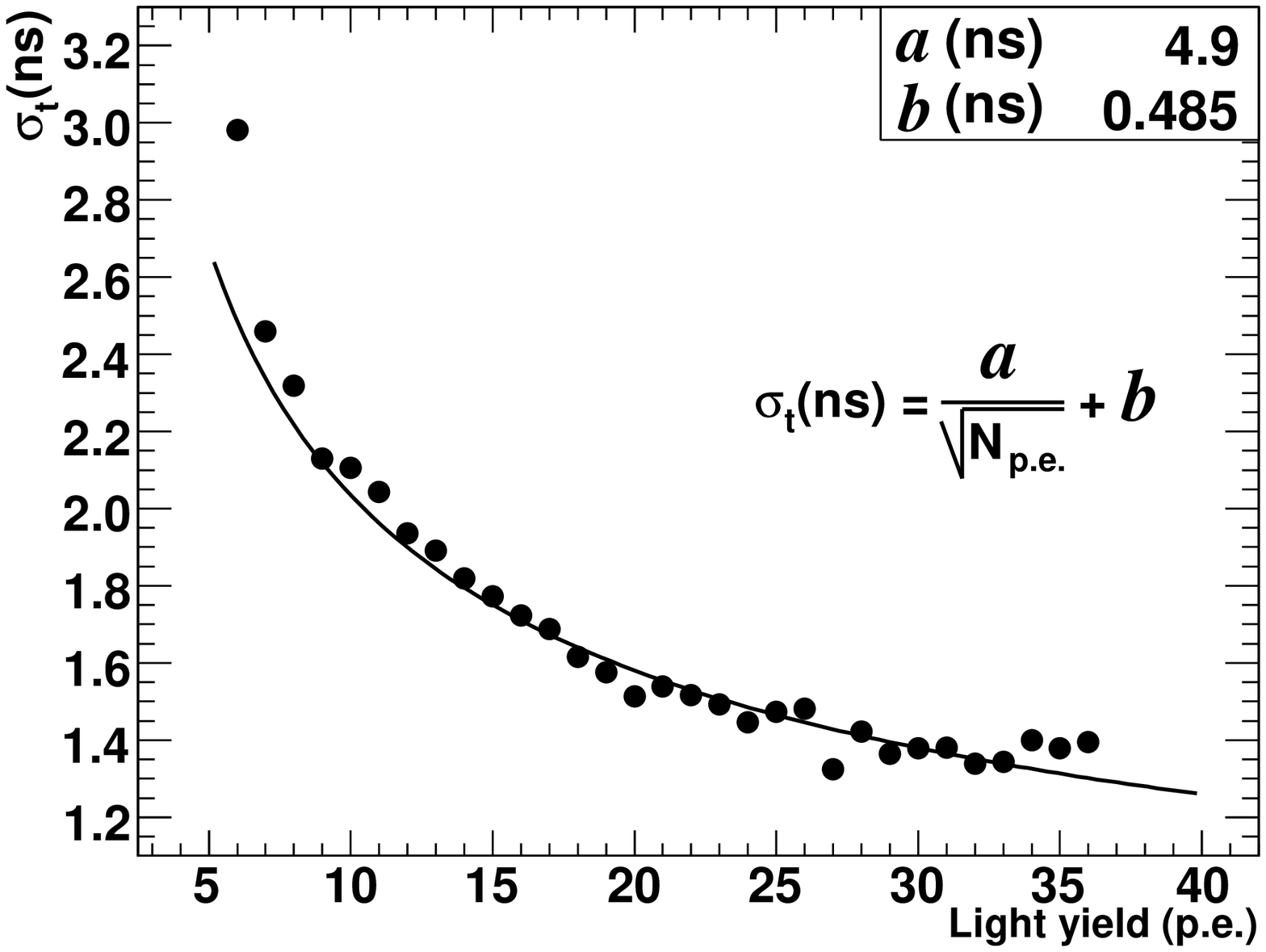}
\caption{The time resolution versus light yield in the center of 
the S--counter.}
\label{fig:timing} 
\end{figure}

\begin{figure}[htbp] 
\centering\includegraphics[width=13cm,angle=0]{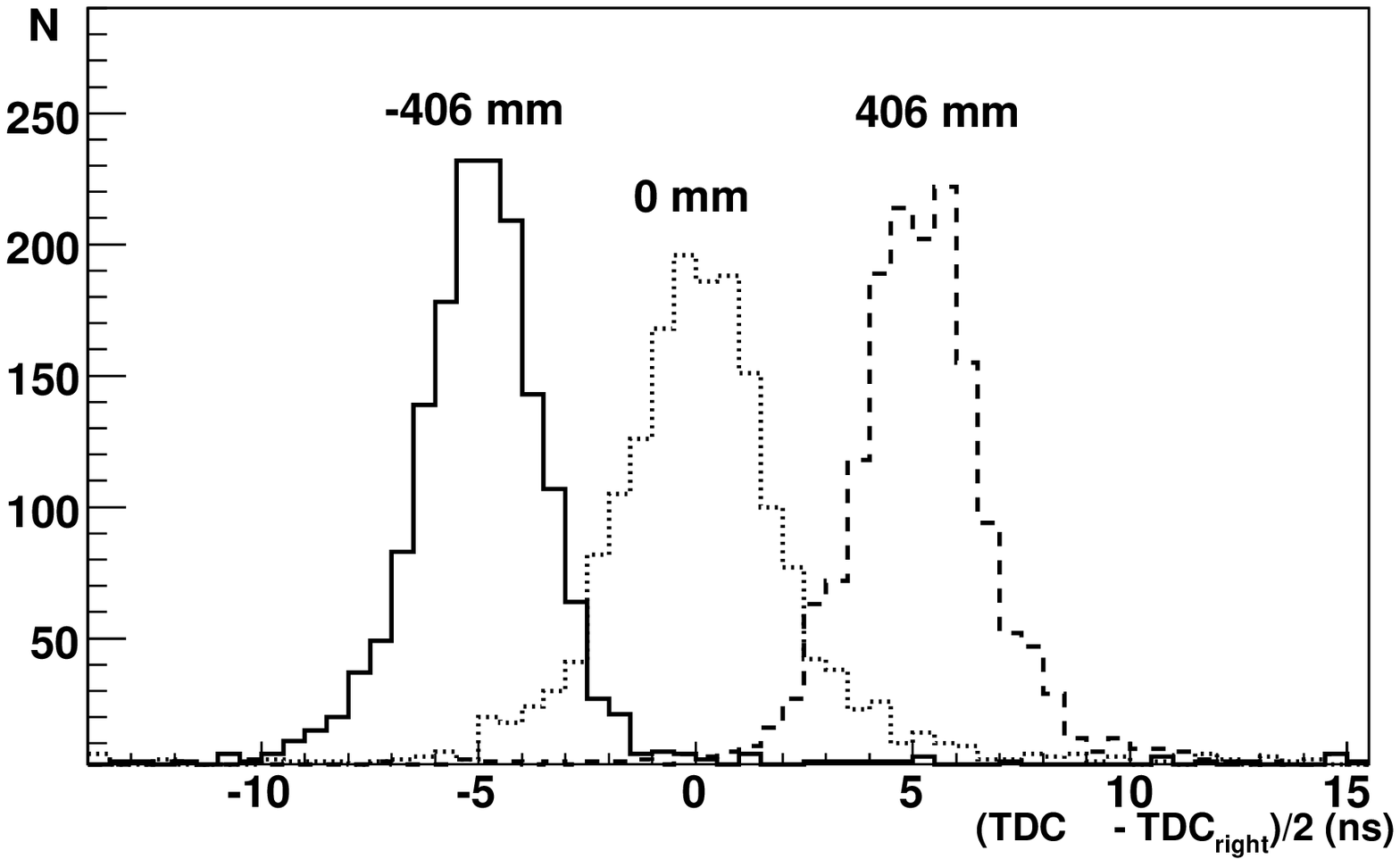}
\caption{Time spectra for 3 positions along the S--counter. A distance of 
81.2~cm corresponds to the  time shift of 11 ns between the left and right 
MRS signals.}
\label{fig:time-shift} 
\end{figure}

The time resolution was measured with 
the discriminator thresholds  set to a  level of 0.5~p.e. for each 
MRS photodiode.  
 To suppress 
the timing spread caused by the trigger  counters (as in the cosmic ray test) 
 we used 
the 
combination $(T_{left} - T_{right})/2$ to determine the 
time resolution. 
The dependence of the  time resolution  on the light yield is presented in 
Fig.~\ref{fig:timing}. 
The time resolution depends on photostatistics and is proportional to 
 $1/\sqrt{N_{pe}}$.  
  At a typical 
l.y. of 15~p.e./MIP $\sigma_t = 1.75$~ns is obtained.  The time resolution of the
S--counter is mainly  determined  by the slow decay time of the Y11 fiber. 
Green light 
travels along a WLS fiber at a speed of 17~cm/ns while the signal propagates 
along the counter at a  smaller speed of 7.4~cm/ns because of the fiber routing. 
The time spectra for 3 positions of the beam along the S--counter are shown in 
Fig.~\ref{fig:time-shift}. 
 The obtained time difference between left and right signals 
 allows us to extract 
the beam coordinate along the S--counter using the expression
\begin{equation}
x(T)=7.4[\frac{\rm cm}{\rm ns}]\times(T_{left}-T_{right})/2.
\label{eq:coordin}
\end{equation}
A spatial resolution of $\sigma_x =13.4$ cm is obtained  in the center of 
the S--counter 
and  $\sigma_{x(T)} = 10.4$ cm near both ends. 

\begin{figure}[htbp]
\centering\includegraphics[width=11cm,angle=0]{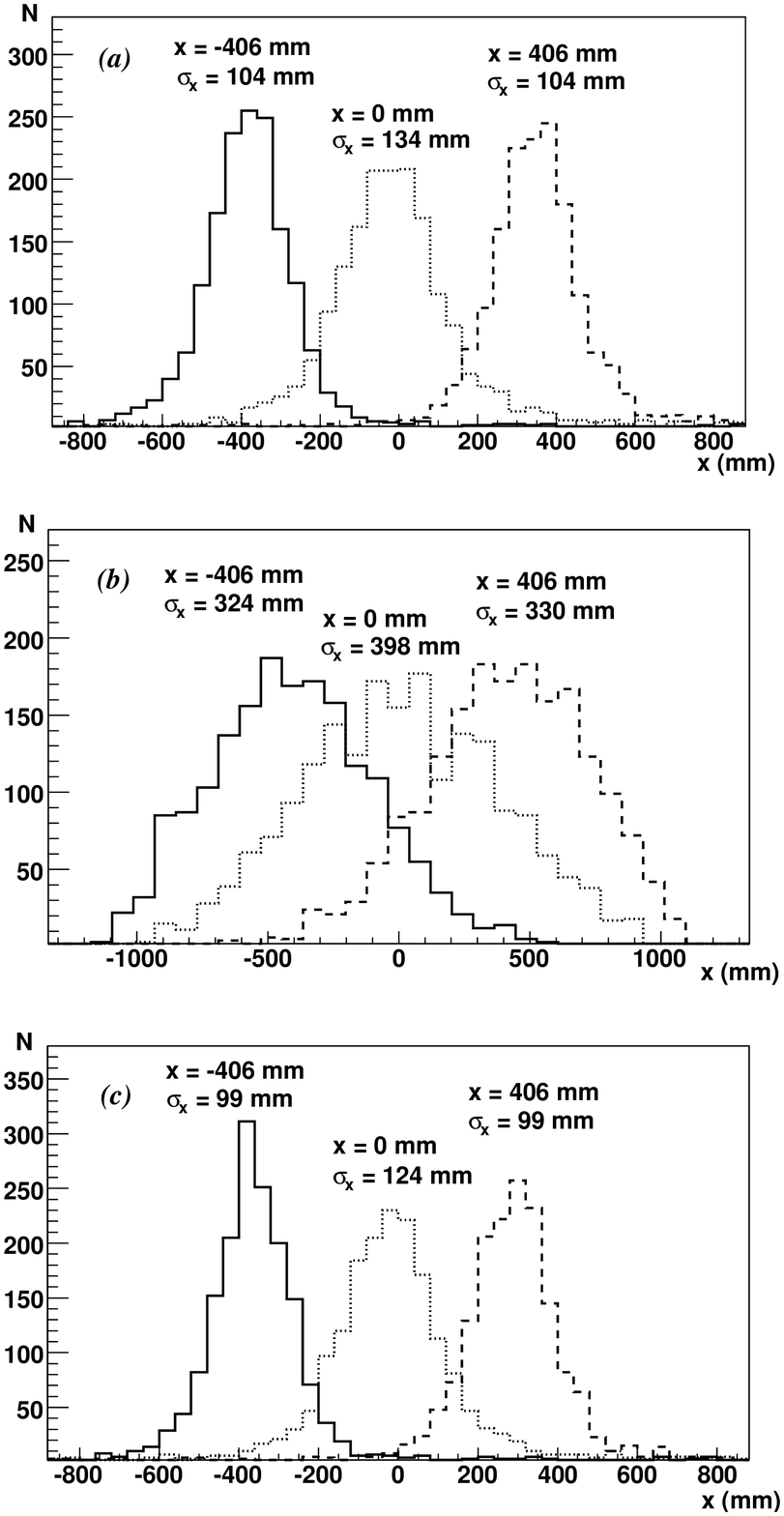}
\caption{The spatial resolution along the S--counter at 3 beam positions: 
(a) the coordinate $x$ is obtained from timing; (b) the coordinate $x$ is 
obtained from the asymmetry  between light yields at the two counter ends; 
(c) the combined position resolution.}
\label{fig:space_resol} 
\end{figure}

The spatial resolution can be improved taking into account the light  attenuation
along the fiber. The  asymmetry between the signals from 
the left and right MRS photodiodes  
$(A_{left} - A_{right})/(A_{left}+A_{right})$ is sensitive  to the hit 
position of a MIP, but the spatial resolution obtained using the l.y. 
attenuation 
is poor ($\sigma_{x(A)} \sim 35 $ cm) because of 
large fluctuations in the light yields. However, the combination of both 
methods 
\begin{equation} 
x=\frac{x(T) + wx(A)}{1+w},
\label{eq:combination}
\end{equation}
 where $x(T)$ and $x(A)$ are the MIP positions obtained 
from timing and amplitude asymmetry with accuracies of $\sigma_{x(T)}$ and 
$\sigma_{x(A)}$, respectively.  The 
weight  $w$ is given by  $w=\sigma_{x(T)}^2/\sigma_{x(A)}^2$,
allows us to slightly improve the spatial
resolution.   Fig.~\ref{fig:space_resol} 
shows the spatial resolution for three  beam positions obtained for the
combination of the signal timing and amplitude. 
The spatial resolution in the center of the S--counter is obtained to be 12.4 cm, 
while it is about 9.9 cm at both ends.  

\section{Conclusion}
The scintillator counters for the SMRD of the T2K near detector have been
designed and tested.
The readout  of the  extruded  scintillator counters is provided via a WLS 
fiber which is embedded into an S--shape groove 
 and viewed from both ends by  multi--pixel avalanche
photodiodes operating in the limited Geiger mode. The studied 
MRS photodiodes demonstrate  good 
performance: a low cross-talk of a few per cent, the photon detection 
efficiency for green light of about 12\%,  and a long 
term stability. These devices are insensitive to magnetic fields, their 
calibration and stability control can be provided by means of the
 excellent p.e.  
peak resolution. The linearity range  of  the tested MRS photodiodes is less 
than 200 p.e. and the recovery 
time is about 5 $\mu$s. Although these parameters might be critical 
for some applications,  this  performance is  acceptable 
for  many detectors 
of the ND280 complex of the T2K experiment.  

An average l.y. of about 15~p.e./MIP, a  MIP detection efficiency 
 greater than 99.5\%,   a time resolution of 
1.75~ns for a MIP, and a spatial resolution of  $\sigma_x = 9.9--12.4$~cm were 
obtained in a pion beam test.

The authors are grateful to D.~Renker, A.~Akindinov and A.~Konaka 
for useful discussions.
This work was supported in part by the ``Neutrino Physics'' Programme 
of the Russian Academy of  Sciences.

 
\end{document}